\documentclass[preprint,12pt,authoryear]{elsarticle}

\usepackage{amsmath}
\usepackage{enumerate}
\usepackage{amssymb}
\usepackage{breqn}
\usepackage{comment}
\usepackage[normalem]{ulem}
\usepackage[usenames,dvipsnames]{color}

\usepackage{setspace} 
\usepackage{lineno}
\journal{Environmental Modeling \& Software}
\begin{document}

\begin{frontmatter}



%%\title{Optimal portfolios for coastal flood risk management: an application of the i-COW framework}

\title{Title\tnoteref{RobTitle}}
\author{Robert L. Ceres\fnref{label2}}
\ead{rlc299@psu.edu}

\title{Title\tnoteref{ChrisTitle}}
%% \tnotetext[ChrisTitle]{Professor of Climate Dynamics}
\author{Chris E. Forest\corref{cor1}\fnref{{label2},{label3},{label4}}}
\ead{ceforest@psu.edu}
\ead[url]{met.psu.edu/people/cef13}
\fntext[label2]{Department of Meteorology and Atmospheric Science, The Pennsylvania State University, University Park, Pennsylvania, USA.}
\fntext[label3]{Department of Geosciences, The Pennsylvania State University, University Park, Pennsylvania
16803, USA.}
\fntext[label4]{Earth and Environmental Science Institute, The Pennsylvania State University, University Park,
Pennsylvania 16803, USA.}

\title{Title\tnoteref{KlausTitle}}
%% \tnotetext[KlausTitle]{Professor}
\author{Klaus Keller\fnref{{label3},{label4}}}
\ead{kzk10@psu.edu}
\ead[url]{geosc.psu.edu/academic-faculty/keller-klaus}

\cortext[cor1]{Corresponding author.}
\address{507 Walker Building\\University Park, PA 16802\fnref{label2}}

\title{Optimization of multiple storm surge risk mitigation strategies for an island City On a Wedge}

\begin{abstract}

Managing coastal flood risks involves selecting a portfolio of different strategies. Analyzing this choice typically requires a model. State-of-the-art coastal risk models provide detailed regional information, but they can be difficult to implement, computationally challenging, and potentially inaccessible. Simple economic damage models are more accessible but may not incorporate important features and thus fail to model risks and trade offs with enough fidelity to support decision making. Here, we develop a new framework to analyze coastal flood risk management. The framework is computationally inexpensive yet incorporates common features of many coastal cities. We apply this framework to an idealized coastal city and assess and optimize two objectives using combinations of risk mitigation strategies against a wide range of future states of the world. We find that optimization using combinations of strategies allows for identification of Pareto optimal strategy combinations that outperform individual strategy options.

\end{abstract}

\begin{keyword}
storm surge\sep flood risk\sep coastal\sep damage\sep resilience\sep multi objective robust decision making\sep risk management\sep New York\sep Hurricane\sep climate



\end{keyword}

\end{frontmatter}


%\begin{linenumbers}
%\doublespacing  
\section{Introduction}\label{into}

Communities have used dikes (or levees)  to protect coastal areas from floods for centuries. In the Netherlands, for example, dikes have been used to protect small regions since the 13\textsuperscript{th} century \citep{gerritsen_what_2005}. The presence of dike rings surrounding low areas was so prevalent that the Dutch language has a word for it, `polder,' which has come into English usage as well \citep{stevenson_oxford_2010}. Numerous other defensive strategies are available to reduce the risk of storm surge. These risk mitigation strategies have advantages and disadvantages impacted, for example, by where the strategies are considered or how they appeal to different stakeholders. These strategies can include insurance, preservation or enhancement of natural barriers, construction of physical barriers  across waterways, installation of active measures such as pumps, adoption of zoning restrictions, withdrawal or relocation of development, physical alteration of buildings, and resiliency improvements \citep{FEMA-P-55,NYC_Mitigation_2014}.

Regardless of which strategy or combination of strategies is considered, policymakers require a means of assessing the needed level of protection and evaluating the performance of candidate strategies in terms of often divergent stakeholder objectives \citep{groves_analysis_2016}. Prior to 1953 in the Netherlands, the predominant practice had been to establish dike height based on the highest previously observed flood levels plus a three foot safety margin \citep{van_dantzig_economic_1956,battjes_coastal_2002}.

Today, a common approach to establishing flood protection levels is based on a return level, the estimated height that is expected to be exceeded for a specified probability of occurrence. In the United States, the 100-year return level is used for this purpose and has led to the widespread use of `base flood' in reference to the expected water level to be reached or exceeded with a 1$\%$ probability in a given year.  The US government uses `base flood' to set policy \citep{ u.s._cfr_725_executive_orders_11988_18_1988, u.s._cfr_725_executive_orders_11988_18_2015, fema_implementing-guidelines-for-eo11988-13690_08oct15_508_2015, bellomo_coastal_1999}. Recently researchers and policymakers have expressed concerns about the adequacy of this standard and the efficacy of current implementation of the standard \citep{highfield_examining_2013, galloway_assessing_2006, wing_estimates_2018}.

Other approaches to estimating flood risks and establishing flood safety measures are possible. In January of 1953 in the Netherlands, a winter storm generated a storm surge, subsequently named the ``Big Flood,'' that caused extensive damage and killed 1,836 people \citep{gerritsen_what_2005}. In response, the government formed the Delta Commission, which turned to the Dutch mathematician David van Dantzig \citep{van_dantzig_economic_1956,gerritsen_what_2005, zevenbergen_taming_2013}. He approached the problem from both a statistical and economic perspective. On the statistical side, he examined historical tide gauge data to estimate the probability that a given flood height had been exceeded. On the economic side, he considered the current cost of constructing a dike and the corresponding net present cost of future damages that should be expected. The sum of these two quantities is the net present cost of the strategy. Minimizing this net present cost results in an economically optimal risk reduction strategy that sets an economically optimal dike height \citep{van_dantzig_economic_1956}.

Researchers have subsequently improved `van Dantzig style' models to account for parameter uncertainty and improve flood probability models \citep{slijkhuis_optimal_1997, speijker_optimal_2000, huisman_hydraulic_2010, oddo_deep_2017, wong_brick_2017} and to investigate methods of learning for optimization of future height adjustments \citep{kok_living_2008,garner_using_2018}. However, several assumptions limit the utility of `van Dantzig style' models. For example, `van Dantzig style' models are restricted to evaluating a single protection strategy, such as construction of a levee or dike. In contrast, policymakers today are interested in considering other approaches to risk mitigation and combinations of different strategies \citep{ligtvoet_climate_2012, new_york_city_special_initiative_for_resilient_rebuilding_planyc:_2013, fischbach_reducing_2017}. A second example of `van Dantzig style' model limitations is their simplifying assumptions that dike cost is proportional to height and damage due to flooding is zero until levees are breached or over topped, at which point 100$\%$ damage occurs \citep{van_dantzig_economic_1956}. These may be reasonable approximations for traditional dikes protecting flat terrain in the low polders found in the Netherlands but are a poor representation of many major cities located on a rising coast or hilly terrain. 

Another potential weakness of a `van Dantzig style' model is its focus on a single objective: the discounted expected total cost. The stakeholder community impacted by flood events and affected by flood risk mitigation efforts is large and diverse, with correspondingly diverse values \citep{harman_global_2015, bessette_building_2017} associated with multiple and often conflicting sets of objectives \citep{harman_global_2015, porthin_multi-criteria_2013, oddo_deep_2017}.

Unfortunately, estimated base flood levels or dike heights do not allow city planners to evaluate changes to damage associated with surge heights above the selected return level. Additionally, return periods for these estimated levels are typically long compared to the record of surge observations available \citep{grinsted_homogeneous_2012,grinsted_projected_2013,menendez_changes_2010,Ceres2017,Lee_Multidecadal}. This relative sparsity of data leads to large uncertainties surrounding estimates of long-period return levels \citep{coles_introduction_2001} and the potential for bias in estimating extreme event risks using common extreme value analysis methods \citep{coles_fully_2003,Ceres2017}.

Researchers have developed more complex methods and models for assessing various aspects of storm surge risk. Superstorms Katrina and Sandy have motivated substantial investments in storm surge protection for both the New Orleans and New York City metropolitan areas, respectively \citep{fischbach_coastal_2012,aerts_low-probability_2013,groves_analysis_2016,aerts_evaluating_2014}. The magnitude of proposed investments justified extensive, site-specific research on both future storm surge risk for these broad geographical regions and for evaluation of several proposals to mitigate that risk. To support these efforts, researchers have developed state-of-the-art storm surge modeling frameworks. These frameworks incorporate hurricane track and intensity prediction models driving hydrological storm surge models, incorporate site-specific bathymetry and geography, that in turn interact with local topography and potential defensive measures to generate simulated inundation levels over a wide area. These inundation levels are then matched to detailed demographic data and damage models to produce area wide and location specific estimates of economic impact \citep[e.g.][]{kerr_u.s._2013, taflanidis_rapid_2013, flowerdew_development_2010,orton_detailed_2012, fischbach_coastal_2012,aerts_evaluating_2014}. These models assist decisionmakers and stakeholders exploring the specific impacts of a storm surge risk mitigation strategy because the models realistically quantify storm surge damages that can be expected for a given set of storms.

In New Orleans, for example, the Coastal Louisiana Risk Assessment (CLARA) model was developed to estimate flood extent, associated flood depths, and resulting damages for the entire Louisiana coastal region \citep{fischbach_coastal_2012,david_r._johnson_estimating_2013}. Policymakers can use this approach to evaluate the effectiveness of several risk mitigation strategies in response to specific storms over the modeled region. For instance, given five storm surge barrier options for Lake Pontchartrain, the CLARA model was recently used to evaluate the reduction in storm surge damage across 15 counties against 77 storms \citep{fischbach_reducing_2017}. In many cases, `CLARA style' frameworks couple several computationally expensive models. For instance, CLARA uses the Advanced Circulation (ADCIRC) and the Simulating Waves Nearshore (SWAN) models to develop water and wave height levels throughout the study region. Computationally, these models have long run times that limit the analysis to relatively few storms and tracks \citep{fischbach_reducing_2017}. Hence, using this approach to evaluate the effectiveness of multiple risk mitigation strategies against the full range of possible future storms can be computationally unfeasible. This, then, influences the selection of optimal mitigation strategies. Additionally, the expense associated with creating similar models for other regions will exceed the resources available to many communities.

Storm surge models of `intermediate complexity' have been developed and can serve an important role in evaluating storm surge risk. As one example, a simplified hydrodynamic model was developed to calculate inundation levels for the Bay of Bengal \citep{lewis_storm_2013}. Another study used a simple statistical model of storm surges coupled to a digital terrain model and gridded demographic information to assess the impact of sea level rise on Copenhagen's risk of catastrophic storm surge \citep{hallegatte_assessing_2011}. A third example utilizes a statistical model of storm surges coupled to a model that simulates the evolution of barrier islands, which, in turn, is linked to an agent based economic model that predicts the performance of resort areas dependent upon those barrier islands \citep{mcnamara_coupled_2008,mcnamara_coupled_2013}. However, these models are location specific, and are limited in their ability to analyze multiple strategies.  As such, the use of intermediate complexity models is infrequent. 

\begin{figure}[h]
    \centering
    \includegraphics[width=5in]{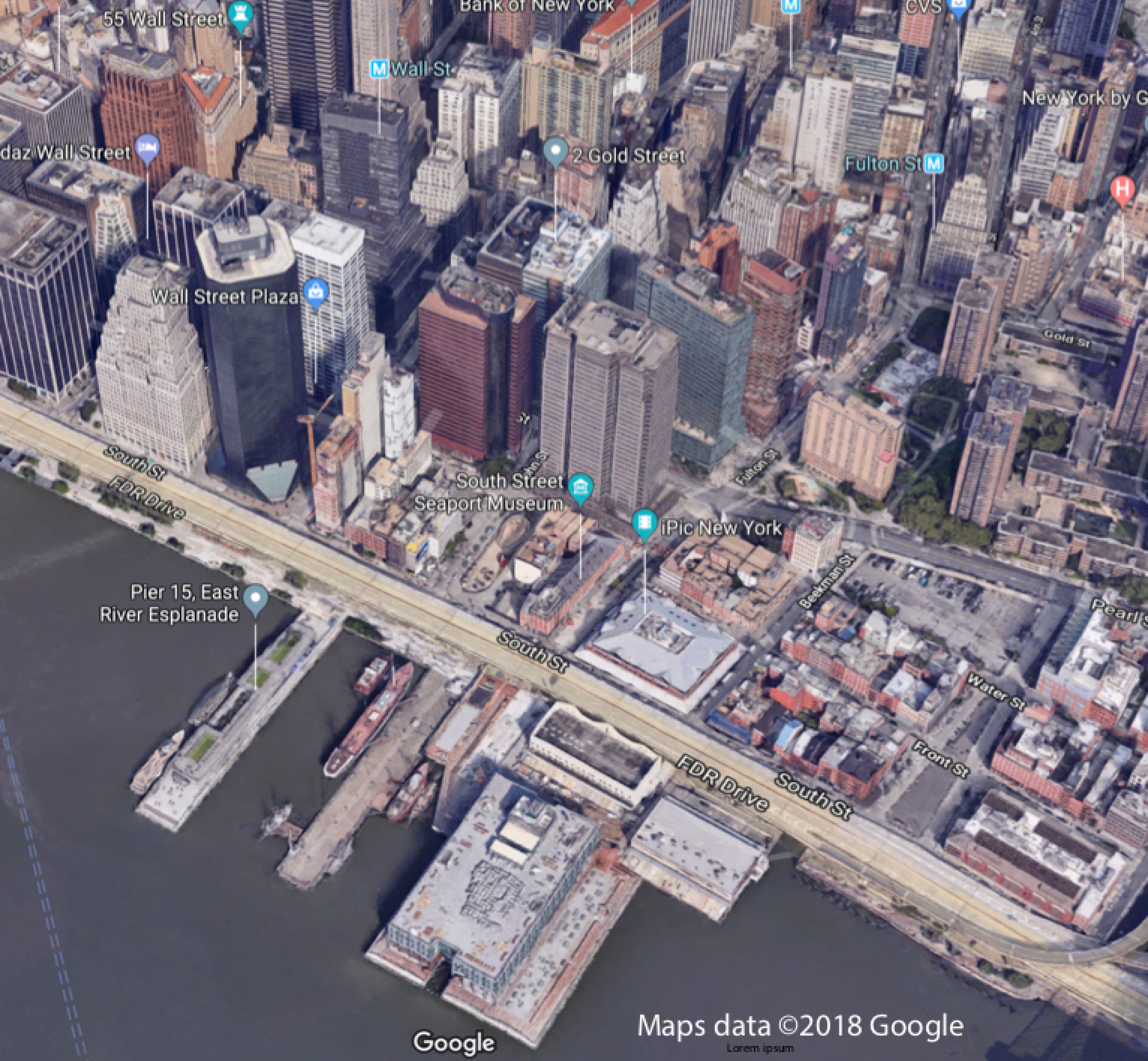}
    \caption[Aerial view of the South Street Seaport area of Manhattan]{An aerial view of the South Street Seaport area of Manhattan, NY that inspired iCOW. The elevation change from South Street to the Fulton subway station (upper right) is approximately 7 meters.  Courtesy Google Maps. (Maps data \textcopyright2018 Google).}
    \label{fig_SouthPort}
\end{figure}

The gaps in storm surge modeling complexity and usage suggests a need for a new framework with different modelling capabilities. Here, we develop the island City On a Wedge (iCOW), a model framework that bridges the gap between `van Dantzig,' and `CLARA style' modeling approaches. Using an idealized geography of a city situated on a rising coastline, such as Manhattan show in Fig. \ref{fig_SouthPort}, the model simulates the increasing damage that occurs with larger storm surges and the distribution of those damages across the city. The iCOW framework can be used to evaluate combinations of multiple risk reduction strategies such as insurance, preservation or enhancement of natural barriers, construction of physical barriers and sea-gates across waterways, installation of active measures such as pumps, adoption of zoning restrictions, withdrawal or relocation of development, physical alteration of buildings, and resiliency improvements. We demonstrate the utility of this model by evaluating combinations of withdrawal, building resistance, and dikes by evaluating several objectives over differing time scales. These objectives can include investment cost, investment timing, median or maximum annual storm surge damage, or the distribution of damage within the city. The framework is flexible enough to incorporate other effects such as economic loss associated with withdrawal strategies or potential value shifts associated with construction of levees or dikes.

%%%
%%%
iCOW requires much less run-time than the CLARA model. As a result, iCOW is capable of evaluating the efficacy of multiple combinations of storm surge risk mitigation strategies over a wide range of potential futures. While computationally inexpensive, iCOW improves upon `van Dantzig style' models in several important respects. The framework incorporates typical characteristic features of coastal cities, features intended to improve its overall fidelity. As a result, the iCOW framework can extend insights gained from `CLARA style' models by evaluating many more combinations of defensive strategies against many potentially divergent objectives, and over a wider range of future risk scenarios. This additional potential insight, however, is limited to a much narrower spatial range (limited to a single community of relatively uniform characteristics) and may provide a less realistic representation of specific future outcomes such as the precise amount of damages expected from a specific storm. The iCOW's lower complexity and more limited spatial extent allows for coupling with Multiple Objective Evolutionary Algorithms (MOEA) to optimize complex combinations of risk mitigation strategies evaluated against multiple and potentially divergent objectives of interest to stakeholder communities. 

Because the iCOW framework is largely self contained and does not demand high resolution geospatial information (such as bathymetry data, topography, ground coverage or utilization), it is considerably easier and less expensive to implement for a particular coastal community compared to `CLARA style' models.

The remainder of this article explains the general characteristics and features of the iCOW framework in terms of an XLRM framework \citep{lempert_general_2006}, describes the iCOW computational environment and experimental design, presents example applications, and discusses results and conclusions.
\section{iCOW XLRM framework}\label{iCOW_features}

\begin{figure}[h]
    \centering
    \includegraphics[width=5.4in]{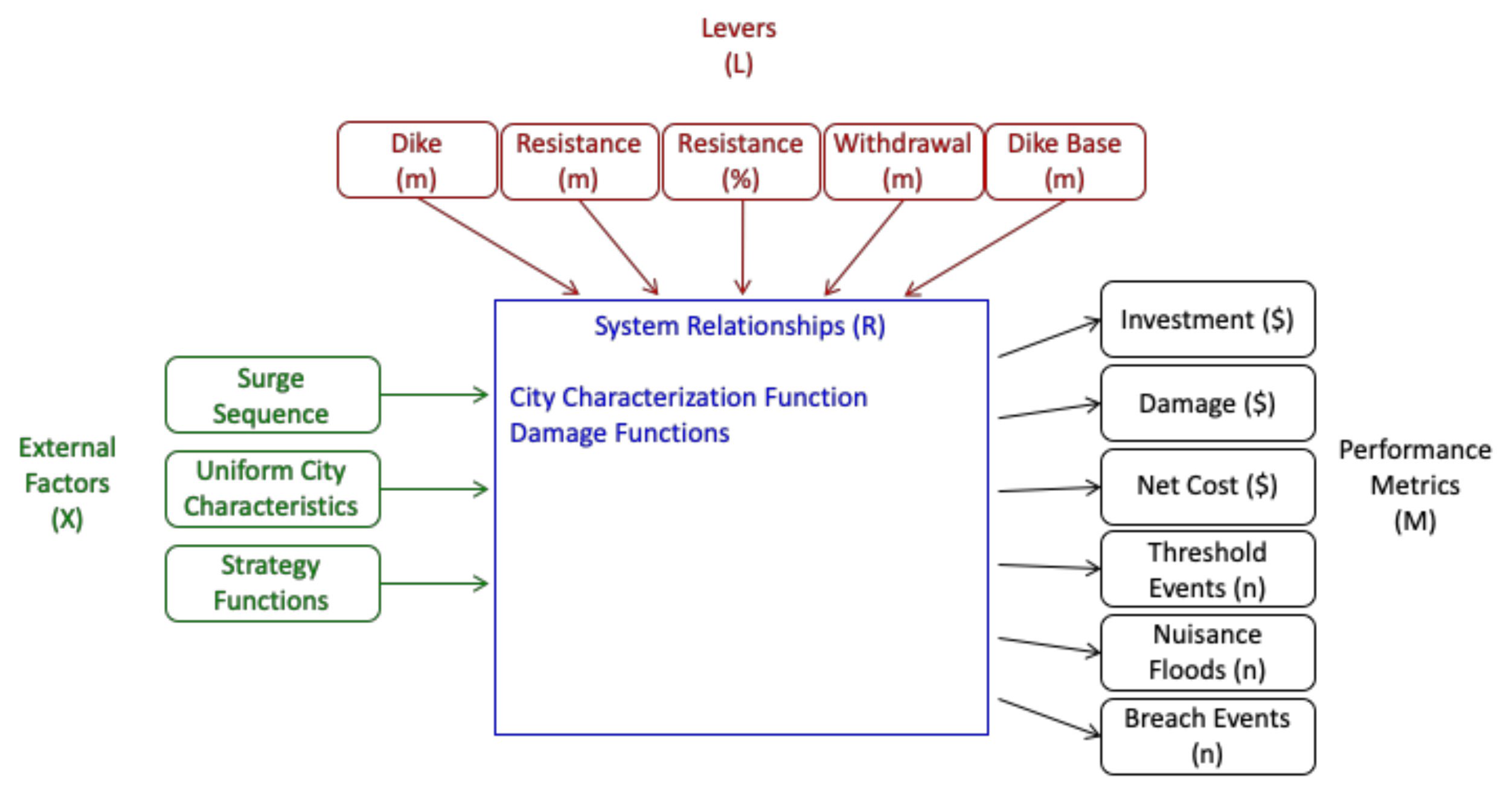}
    \caption[ICOW XLRM diagram]{Top level iCOW model XLRM diagram. External factors (X) in green, levers (L) in red, Relationships (R) in blue and performance metrics (M) in black. Investment cost, damage, and net cost are considered in this article, but other metrics, such as the incidence of large threshold events, nuisance events, or breach events are possible.} 
    \label{figXLRM}
\end{figure}

We describe the iCOW framework using the XLRM framework for robust decision making \citep{lempert_general_2006}. Fig.~\ref{figXLRM} shows the logical relationships between exogenous factors (X), model strategy levers (L), modeling relationships (R), and performance metrics (M). Exogenous factors (X) are characteristics of the simulated city environment that are fixed from the perspective of decisionmakers and stakeholders. Levers (L) simulate the actions that decisionmakers can implement to affect the city's response to storm surge. Metrics (M) define outcomes that are of potential interest to different stakeholders. Relationships (R) represent the model logic implemented to simulate the city response in terms of metrics that result from a particular combination of exogenous factors and lever settings.
%%%
%%%

For this article, and to demonstrate the capabilities of the iCOW framework in general, we use many simplifying assumptions regarding the modeled city's characteristics. The iCOW design and associated parameterizations are intended to represent the general qualities and behaviors of the XLRM components described below. Peer reviewed literature and quantified data for these categories are often not available (and require justification as discussed below.) Additionally, location specific information relevant to XLRM categories (e.g., Dike construction costs) often varies substantially from location to location. Similarly, the simplified topography and demographics of the modeled city (e.g., the regular wedge shape geometry of the modeled community) are intended to be representative of a generic coastal community rather than any particular site. iCOW framework results are sensitive to these simplifying assumptions and representative and generic parameter settings.

\subsection{Exogenous factors}
 
 Within the XLRM framework \citep{lempert_general_2006}, exogenous factors represent any condition that has an impact on the output metrics that cannot be changed by decisionmakers or stakeholders. For this article, the primary exogenous factors are the initial city parameters and the sequence of storms that will impact the city. We incorporate other exogenous factors as model parameters (e.g., the per volume cost to construct a dike or the height of the city seawall) (see supplemental table \ref{parameter tables} for full details). 

\subsubsection{Surge simulation}\label{surge generation module}
Storm surges are simulated as annual highest surge heights generated from a nonstationary generalized extreme value (GEV) model such that the 100-year storm surge is increasing at a rate of one meter per century due to an increase in both the location and scale parameters \citep{Ceres2017}. Some storm surges generated in this manner may be larger than physically plausible. Storm surges used for this study are clipped such that surge heights exceeding a threshold are capped at the threshold. We use a thresholds of 12 m for the first year and increase by 0.01 m per year thereafter. This threshold is chosen as a compromise between the goal of accounting for the full range of risk exposure based on the statistical model used to generate the surges and the desire to provide fidelity to physical reality.

We use 5,000 realizations of 50-year sequences of storm surges for all examples discussed in this article. This number represents a compromise between computational efficiency and stable results. See the supplemental materials, (section \ref{futures}), for additional figures and discussion on the number of realizations used.

\subsubsection{Initial city characteristics}\label{city characteristics}

Many cities (e.g. Boston, NYC, or San Diego), are situated at the water's edge and consist of a gradually rising terrain. In these cases, higher surge levels will result in larger areas and greater depths of inundation. The waterfront areas in these cities (such as Manhattan, New York City, NY, Fig.~\ref{fig_SouthPort}) are often densely packed with tall buildings. iCOW simulates an island city situated on a rectangular wedge. The most prominent characteristic of the iCOW city is the gradually rising elevation of the city with distance from the city's lowest waterfront (Fig.~\ref{fig-iCOWSchematic}). The city terrain is elevated from the normal water level by a seawall. iCOW buildings are uniformly tall and higher than the highest potential surges.

Damage to buildings accrues based on the volume of the building flooded. Buildings have a basement volume that floods completely when water reaches the level of the building. Based on volume, real estate value is initially assumed to be constant regardless of building height or location relative to the waterfront. In the simplest configuration, city value and density are uniformly and continuously distributed from water's edge to the highest city elevation. As defensive strategies are added, this value density may change. The addition of defensive strategies divides the city into zones of different damage vulnerabilities as described below. 

%%\subsection{Risk mitigation strategies and city zones}\label{strategies and zones}
\subsection{Strategies and Levers}\label{strategiesAndLevers}

\begin{figure}[h]
    \centering
    \includegraphics[width=5.4in]{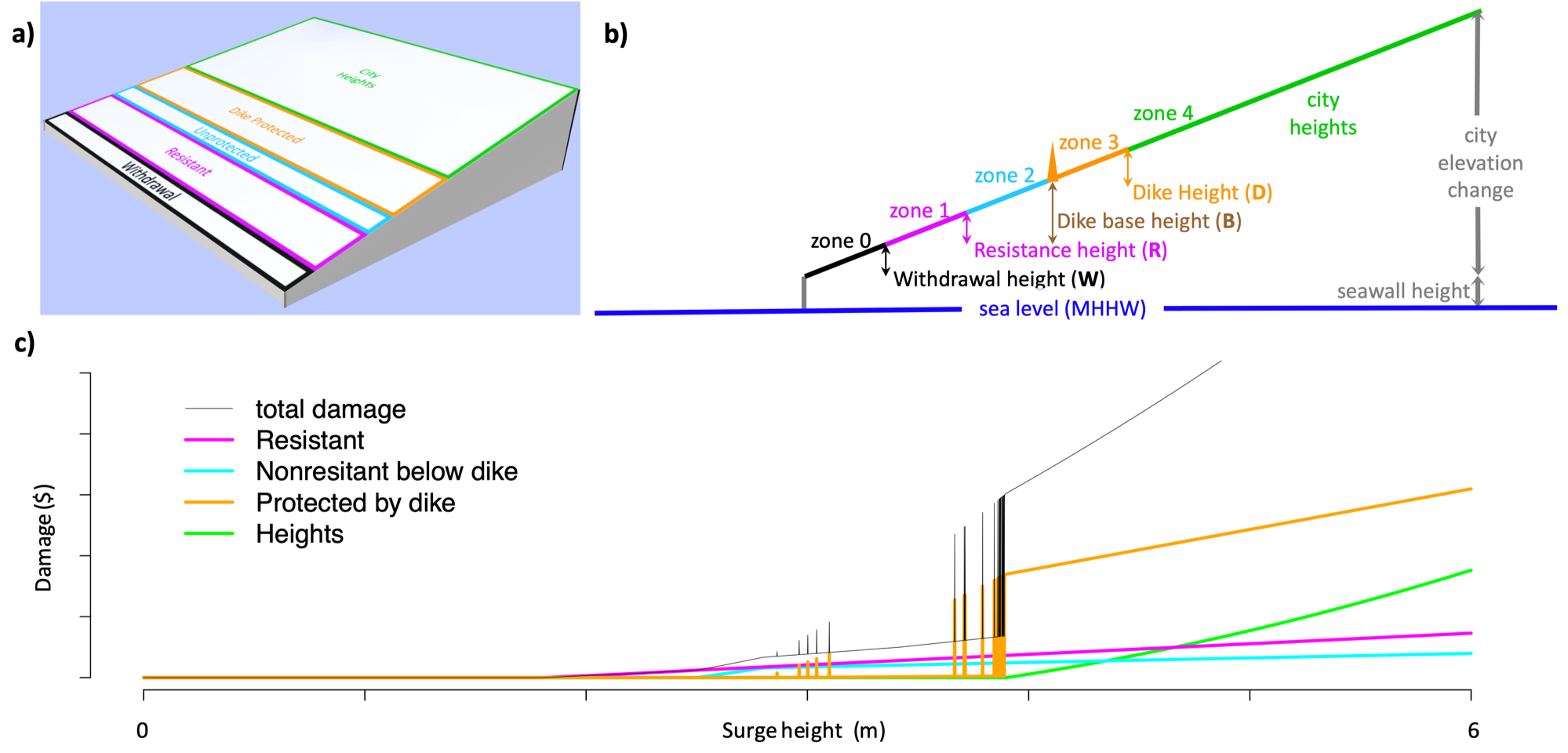}
    \caption[iCOW topology and damage profiles]{The overall topology of the iCOW model is shown in panels a and b. In Zone 0 (black) city infrastructure is relocated to higher elevations, thus no damage accrues. In Zone 1 (magenta), buildings are resistant to surge damage to a specified elevation. If this elevation is lower than the dike base height an unprotected area, Zone 2 (grey) exists. Zone 3 (orange) lies between the dike base height and the dike height and is protected by the dike. Damage will only occur if the dike fails or is over topped. Zone 4 (green), the city heights, extends from the dike top elevation to the highest area of the city and is unprotected. Damage curves for each city zone (by color) and the total city damage (black) are shown in panel c. Curves are generated using iCOW to evaluate a combined strategy against a range of surge heights. In the dike protected area (zone 3 in orange), damage accrues at a very low rate for surges below the dike height unless the dike fails or is overtopped. Instances of dike failure are indicated by the damage spikes in this section of the figure.}
    \label{fig-iCOWSchematic}
\end{figure}

As discussed in the introduction, numerous defensive strategies are available to mitigate storm surge risk. For this study, we consider one fixed and three adjustable defensive strategies that can be implemented to varying degrees and in combination. The fixed defensive strategy is the presence of a seawall around the city. The three adjustable defensive strategies are: i) withdrawal from at-risk areas, ii) improving resistance to damage, and iii) construction of a dike. All strategies are assumed to be implemented uniformly across the width of the city parallel to the coastline (see Fig. \ref{fig-iCOWSchematic} a). Within iCOW, levers are the mechanisms by which defensive strategies are implemented.

%%%
%%%

%%%
%\begin{figure}[h]
%    \centering
%    \includegraphics[width=5in]{iCOW_Damage.pdf}
%    \caption[iCOW topology and damage profiles]{The overall topology of the iCOW model is shown panel a. In Zone 0 (black) city infrastructure is relocated to higher elevations, thus no damage accrues. In Zone 1 (magenta) buildings are resistant to surge damage to a specified elevation. If this elevation is lower than the dike base height an unprotected area, Zone 2 (grey) exists. Zone 3 (orange) between the dike base height and the dike height is protected by the dike. Damage will only occur if the dike fails or is over topped. Zone 4 (green), the city heights, extends from the dike top elevation to the highest area of the city and is unprotected. Damage curves for each city zone (by color) and the total city damage (black) are shown in panel b. Curves are generated using iCOW to evaluate a combined strategy against a range of surge heights. In the dike protected area (zone 3 in orange) damage accrues at a very low rate for surges below the dike height unless the dike fails or is overtopped. Instances of dike failure are indicated by the damage spikes in this section of the figure.}
%    \label{fig-iCOWSchematic}
%\end{figure}
%%%

\subsubsection{Seawall}
The iCOW design assumes an existent seawall of fixed height surrounding the city. No damage occurs to the city as long as surge heights are below the level of the seawall. When surge height exceeds the seawall, the height of water that affects the city is equal to the amount of excess and an additional height caused by wave run-up. For this study, there is no option to increase the seawall height, thus iCOW does not include seawall height as a lever.

\subsubsection{Withdrawal}
The first iCOW adjustable strategy is to specify withdrawal from regions of the city below some withdrawal height and relocation to the city's higher levels. This height is the single withdrawal strategy lever. An actual withdrawal strategy implemented in a city such as New York might involve relocating individual buildings at very low elevations that are the least resistant to flood damage. In the model, all unmodified buildings have the same resistance to flooding damage, thus the most vulnerable buildings are uniformly located at the lowest elevations. In the modeled withdrawal strategy, it is this lowest fractions of building volume (without regard to individual buildings) that is withdrawn as withdrawal height increases. This modeled representation seems reasonable so long as the area of the city relocated (the withdrawal height) remains low. As the withdrawal height increases, the fidelity of the modeled withdrawal strategy to an actually implemented strategy would be expected to decrease. In iCOW, we model this relocation of the lowest elevation buildings by creating a zero value zone (zone 0) and redistributing the value that had been contained in that zone over the remaining area of the city, subsequently and uniformly increasing the value density in the remaining city area. 

\subsubsection{Resistance}
The second adjustable strategy is implementation of resistance to damage for the lowest portions of buildings, which are most exposed to flooding. There are two levers associated with this strategy. The first lever, resistance height, \textbf{R}, is a specified height above the withdrawal height. Below this level, buildings are modified to increase their resistance to flood damage. Implementing resistance creates a resistant zone (zone 1). The upper extent of zone 1, in terms of the city elevation, is constrained by the lower of the resistance height or by the location of a dike base. Building volume flooded below the building's resistance height is subjected to less damage than nonresistant buildings. The percentage \textbf{P} of resistance to damage (compared to the non-resistant damage) that results in resistant sections is the second lever resistance strategy lever.  Flooded volumes above the resistance height are subjected to damage at the nonresistant rate. 

\subsubsection{Dikes}
Construction of a dike is the third strategy considered for this study. As with implementation of the other strategies, we assume dikes are constructed parallel to the shoreline. Because iCOW is modeled as a wedge shaped island, the dike must have sections on either end of the dike to keep surges from flowing around the sides of the dike. iCOW dikes greatly reduce flooding damage to buildings located behind them unless they fail or are overtopped (discussed further in section \ref{DikeRelationships}). Two levers implement the dike strategies. The first lever is the elevation of the dike base, \textbf{B}, above the withdrawal height. The second lever is the height of the dike, \textbf{D}. The area behind (above) the dike base but below the dike top are protected by the dike and define zone 3. 

\subsubsection{Strategy combinations}
iCOW can simulate simultaneous implementation of these three strategies through the simultaneous operation of the five model levers. The levers are summarized in Table \ref{levers}.

\begin{table}[h]
\centering
\caption{Levers}
\label{levers}
\begin{tabular}{|r|l|l|}
\hline
Variable   & Name                        & Description \\ \hline
\textbf{W}  & \begin{tabular}[c]{@{}l@{}}\textbf{W}ithdrawal\\height\end{tabular}  & \begin{tabular}[c]{@{}l@{}}height from the seawall below which all\\ structures are relocated  \end{tabular} \\ \hline
\textbf{R}  & \begin{tabular}[c]{@{}l@{}}\textbf{R}esistance\\height\end{tabular}  & \begin{tabular}[c]{@{}l@{}}height from W that structures are resistant to\\damage                 \end{tabular} \\ \hline
\textbf{P}  & \begin{tabular}[c]{@{}l@{}}resistance\\\textbf{P}ercent\end{tabular} & \begin{tabular}[c]{@{}l@{}}percentage reduction achieved in resistant city\\volume             \end{tabular} \\ \hline
\textbf{D}  & \begin{tabular}[c]{@{}l@{}}\textbf{D}ike\\height\end{tabular}        & \begin{tabular}[c]{@{}l@{}}height of a dike measured from the dike base \end{tabular} \\ \hline
\textbf{B}  & \begin{tabular}[c]{@{}l@{}}dike\\\textbf{B}ase\end{tabular}          & \begin{tabular}[c]{@{}l@{}}elevation of the dike base measured from W \end{tabular} \\ \hline
\end{tabular}
\end{table}

The relationship between strategy levers and resultant zones is illustrated in Figs.~\ref{fig-iCOWSchematic}a and \ref{fig-iCOWSchematic}b. 

\subsection{Metrics}
Implementing storm surge mitigation strategies has many consequences, and the importance of these consequences may vary among different stakeholders. To the extent that these consequences can be measured, there are many possible metrics associated with risk mitigation performance. As examples, these could include implementation costs, maintenance costs, the cost to repair buildings damaged by flood, the direct or indirect economic damage associated with storm-driven changes to the economy, damage to important city infrastructure, probabilities of nuisance flooding, the probability of catastrophic events, or even deaths. Some consequences may be more difficult to model, such as a strategy's impact on a community's city character and culture. iCOW is able to calculate many metrics corresponding to the objectives of many different stakeholders. To simplify the illustration of the impact of allowing for multiple strategies and levers, we examine three metrics: implementation cost, expected damage costs, and the total cost, which is the sum of investment cost and expected damage cost.

\subsection{Relationships}\label{relationships}
Within an XLRM modeling framework, relationships convert exogenous factors and lever settings into metrics. In this section, we discuss the relationships between exogenous factors, lever settings, and the cost and damage metrics for each strategy.

In general, iCOW model metrics, including cost and damage, are based on volume of the city effected. In the most basic case, where no strategies are implemented, cost is zero and damage will be proportional to the value density of the area flooded and the building volume flooded.

Actual building damage from storm surges in general increases with exposure to higher water levels but is also affected by other factors such as wave action, flow velocity, and duration of inundation \citep{FEMA-P-55,fema_implementing-guidelines-for-eo11988-13690_08oct15_508_2015}. For simplicity, we model damage based on storm surge height. The actual damage function for particular buildings, however, varies by building usage, construction type, and more broadly, by regional location \citep{prahl_damage_2018}. The iCOW framework adopts a generic damage model where, in an unprotected state, damage to a structure starts occurring when the water level exceeds the building base elevation. At that level, a set damage occurs associated, for instance, with basement flooding. Above this level, additional damage is proportional to the volume of the building flooded.  When calculated in this way, the damage function, which emerges for aggregated areas of the iCOW city, will align with other research into damage functions for urban areas \citep{prahl_damage_2018}. This basic relationship is modified by the employment of strategy levers as discussed below.

\subsubsection{Withdrawal costs and damage relationships}

Conceptually, for a given building, iCOW withdrawal cost can be thought of as the total cost to acquire a new building at a higher location, the cost to relocate, and the cost to remove the old structure. The cost to implement a withdrawal strategy is based on the area to be relocated, the value density of that area, and the total remaining area in the city available for relocation. A fraction of the displaced infrastructure will relocate outside the city. The cost to implement withdrawal, $C_w$, is 
\begin{equation}
\label{costToImplementWithdrawal}
C_w =\frac{v_i*\textbf{W}*f_w}{city\ height-\textbf{W}},
\end{equation}

\noindent where $v_i$ is the initial city value, and $f_w$ is a factor intended to adjust for any local conditions (such as the presence of historic buildings, or heavy industry facilities), as well as relocation and demolition costs that might make it more or less expensive to relocate.

The value of the city after withdrawal, $v_w$, is
\begin{equation}
\label{valueCityAfterWithdrawal}
v_w = v_i*\left(1-\frac{f_l*\textbf{W}}{city\ height}\right),
\end{equation}

\noindent where $f_l$ is the faction of infrastructure that will leave rather than relocate within the city.

Once withdrawal is implemented, no damage will occur to city areas below the withdrawal height, \textbf{W} (zone 0). Above this height, city density is higher and damage will therefore be proportionally higher per volume when surge heights reach the levels above the withdrawal area. 

\subsubsection{Resistance cost and damage relationships}

Modifying buildings to be completely invulnerable to storm surge damage is generally infeasible due to increasing costs, and costs vary based on building type and materials. Intuitively, the cost of implementing resistance should rise as the percentage of resistance, \textbf{P}, incorporated increases, and this cost increases sharply as resistance approaches 100\%. We model this characteristic cost structure using a linearly increasing resistance cost ($c_R$) per unit value until a threshold percentage is reached. As resistance percentage increases above this value, we add an exponentially increasing term that increases sharply as resistance percentage approaches 100\%, so that the cost fraction of resistance with respect to P ($f_{c_R}$) increases  according to

\begin{equation}
\label{resistancePercentEqution}
f_{c_R} = f_{lin}*\textbf{P}+\frac{f_{exp}*max(0,\textbf{P}-threshold)}{(1-\textbf{P})},
\end{equation}

\noindent where $f_{lin}$ is a factor that controls the linear rate of increase in cost at low percentage increases and $f_{exp}$ is a factor that controls the exponential rate of increase.

At low percentages, increasing the percentage resistance to damage results in a linear increase in cost. As resistance percentage increases above a threshold and approaches one (where building volume would be completely invulnerable to surge damage), cost per volume increases sharply such that increasing resistance fraction to one would be infinitely costly.

We assume that the total cost of implementing resistance increases at this rate in proportion to the volume being made resistant, but because the width and slope of the city is constant, the relationship can be greatly simplified and expressed solely in terms of elevation changes. In cases where resistance is not constrained by the presence of a dike, resistance cost is 

\begin{equation}
\label{resistance cost equation not constrained}
c_R=\frac{v_w*f_{c_R}*\textbf{R}*(\textbf{R}/2+b)}{h*(city\ elevation-\textbf{W})},
\end{equation}

\noindent where $b$ is the representative basement depth, and $h$ is the the city building height. In this case, there will be an unprotected zone 2 in front of the dike.

When the resistance zone 1 is constrained by a dike (i.e. \textbf{R} is higher than \textbf{B},) then there is no zone 2 and the areas behind the dike are in the area protected by the dike, zone 3. No resistance is incorporated behind the dike regardless of the resistance height \textbf{R}. When zone 1 is constrained by the presence of a dike, the resistance cost is given by

\begin{equation}
\label{resistance cost equation constrained}
c_R=\frac{v_w*f_{c_R}*\textbf{B}*(\textbf{R}-\textbf{B}/2+b)}{h*(city\ elevation-\textbf{W})}.
\end{equation}

\noindent In the resistant area (zone 1), damage is reduced by the resistance fraction for resistant volumes flooded. When flooding exceeds the resistance height, additional damage accrues at the normal, nonresistant rate for volume flooded above the resistant height.

\subsubsection{Dike cost and damage relationships}\label{DikeRelationships}

Empirical data and peer reviewed literature on dike costs is sparse and indicates a wide range of dike construction costs \citep{jonkman_costs_2013}, thus we make several simplifying assumptions with the aim of modeling the generic case that could be modified as needed to match the particular circumstances of an actual coastal community. Real dikes have sloped sides, or, when they have a constant width profile with respect to height, \textbf{D}, require greater strength at the base compared to the top of the dike. Therefore, with all other factors being equal, taller dikes are more expensive than shorter ones \citep{zhu_up_2009,jonkman_costs_2013}. iCOW currently models all surge barriers as having sloped sides. iCOW dikes have sloped sides, and a flat top, thus the volume of a dike increases approximately geometrically with height. Dikes are modeled as perpendicular to the waterfront. Because the city is modeled as a wedge shaped island, and to prevent surges from flowing around the edges of the dikes, iCOW dikes are U shaped. Therefore, the overall length of a dike will increase, based on the additional length of dike required to be constructed on the sloping sides of the city. These portions of the dike are irregular tetrahedrons. Because the city slope, S, is low, the length of the wings is long compared to the dike height which allows for the simplifying assumption that dike wing lengths at the top and bottom of the dikes are equal. iCOW modeled dike cost is proportional to volume. Additionally, large scale projects such as dikes typically incur a fixed startup cost, which reflects the costs necessary to plan, design, and approve projects, cost to acquire and prepare dike sites, and to make the initial and wrap up costs for large scale projects that are dependent on site location and dike height or volume \citep{zhu_up_2009, jonkman_costs_2013}. For simplicity, we emulate start up cost as a fixed additional height. 

Dike volume ($V_d$) used to calculate cost is based on the a height $h$ (which is the sum of the design height and a fixed initial startup height), city width, $W_{city}$, dike top width, $w_d$, slope of the dike sides, $s$, and slope of the city, $S$, according to

%%
%%\begin{equation}
\begin{multline}
\label{dike volume}
V_d=W_{city}h\left(w_d+\frac{h}{s^2}\right)+\frac{1}{6}\Bigg[-\frac{h^4\left(h+\frac{1}{s}\right)^2}{s^2}-\frac{2h^5\left(h+\frac{1}{s}\right)}{S^4}-\frac{4h^6}{s^2S^4}+\\ \frac{4h^4\left(2h\left(h+\frac{1}{s}\right)-\frac{4h^2}{s^2}+\frac{h^2}{s^2}\right)}{s^2S^2}+\frac{2h^3\left(h+\frac{1}{s}\right)}{S^2}\Bigg]^{\frac{1}{2}}+w_d\frac{h^2}{S^2}.
\end{multline}
%%\end{equation}

\noindent As a result, startup costs will be larger but account for a lower percentage of total costs for taller dikes. The iCOW model specifies dikes based on the location of the dike base relative to the withdrawal height and the height of the dike from the dike base. Cost of the dike, $c_D$, is calculated by multiplying dike volume, $V_d$ from equation \eqref{dike volume} with the per cubic meter cost of the dike, $c_{dpv}$, to yield

%%equation
\begin{equation}
\label{dike cost}
c_D=V_d*c_{dpv}.
\end{equation}

\noindent
The resultant dike height to cost profile is shown in supplemental Fig. \ref{fig_van_Dantzig}.

When dikes function properly and when they are not overtopped, they greatly reduce the damage in the protected zones. Dikes, however, will fail when they are overtopped, and they may fail for a variety of reasons  prior to water levels exceeding their design protection heights \citep{tobin_levee_1995,apel_flood_2004,sills_overview_2008}. Example reasons for failure include: improper design, incorrect operation, inadequate maintenance, or foundation erosion. iCOW represents this nonzero probability of failure with a low fixed probability of failure at all surge heights ($h_{surge}$) below a high percentage of the dike height. Above this threshold, $t_{df}$, probability of failure increases linearly with height until it reaches one at the dike's design height, when Surge height is less than the threshold,

%%equation
\begin{equation}
\label{dike failure}
p_{df}=\frac{h_{surge}-t_{df}}{\textbf{D}-t_{df}}.
\end{equation}

\noindent When the dike holds, damage in zone 3 is reduced by a fraction of what would otherwise accrue in the dike protected zone. If the dike fails or is overtopped, damage will accrue in the dike protected zone at a rate greater than the unprotected rate.

\subsubsection{Aggregate damages}
Damage accumulating by zone for a range of surge heights for one combination of mitigation strategies is illustrated in Fig.~\ref{fig-iCOWSchematic}b. Damage to each zone ($d_Z$) is calculated based on the value of the zone $Val_Z$, the volume of the zone, $Vol_Z$, the volume flooded, $Vol_F$, and the per volume rate of damage incurred by flooding based on the strategy implemented in the zone, $f_{damage}$,
\begin{equation}
\label{zone damage}
d_Z=Val_Z*\frac{Vol_F}{Vol_Z}*f_{damage},
\end{equation}

\noindent
where $Vol_F$ depends on both levers and surge height.

iCOW features many permutations of this damage function depending upon the lever settings, the resulting city zone configuration, and the height of the surge relative to the heights of the city zones and implemented protective strategies.

The iCOW model parameters used in this paper are inspired by the situation in Manhattan in that the building heights are tall relative to potential storm surges, dikes are inexpensive relative to the assets protected behind them, there is an existing seawall, and the breadth of the coastline is long relative to the volume of city protected. The model parameters are easily customized such that they can represent a particular city with greater fidelity. Differences between iCOW and `van Dantzig style' models are summarized in Table \ref{van Dantzig comparison}. At this point, we have presented the iCOW setup and definitions. The implementation is described in section \ref{iCOW framework}, results are described in section \ref{results}, and conclusions are summarized in section \ref{conclusions}.

\begin{table}
\centering
\caption[iCOW van Dantzig comparison]{A comparison of `van Dantzig style' model and the iCOW framework.}
\label{van Dantzig comparison}
\begin{tabular}{l|l|l|}
\cline{2-3}
                                            & `van Dantzig style'                                                            & iCOW model framework                                                                                                                                          \\ \hline
\multicolumn{1}{|l|}{\begin{tabular}[c]{@{}l@{}}area\\ protected\end{tabular}} & polder                                                                & zones based on protection strategies                                                                                                          \\ \hline
\multicolumn{1}{|l|}{\begin{tabular}[c]{@{}l@{}}protection\\ strategies\end{tabular}} & dike                                                                  & withdrawal, resistance, dike                                                                                                                  \\ \hline
\multicolumn{1}{|l|}{\begin{tabular}[c]{@{}l@{}}damage\\ functions\end{tabular}}      & \begin{tabular}[c]{@{}l@{}}fixed cost upon\\ dike topping\end{tabular}& \begin{tabular}[c]{@{}l@{}}cost proportional to volume\\ flooded, taking into account, the\\ protective strategies implemented.\end{tabular}   \\ \hline
\multicolumn{1}{|l|}{\begin{tabular}[c]{@{}l@{}}dike\\ cost\end{tabular}}             & \begin{tabular}[c]{@{}l@{}}proportional to\\ height\end{tabular}      & \begin{tabular}[c]{@{}l@{}}proportional to dike volume with \\ start-up cost\end{tabular}                                                     \\ \hline
\multicolumn{1}{|l|}{\begin{tabular}[c]{@{}l@{}}resistance\\ cost\end{tabular}}       &  n/a                                                                  & \begin{tabular}[c]{@{}l@{}}proportional to volume of city\\ protected and dependent on the\\ fraction of resistance implemented.\end{tabular} \\ \hline
\multicolumn{1}{|l|}{\begin{tabular}[c]{@{}l@{}}withdrawal\\ cost\end{tabular}}       &  n/a                                                                  & \begin{tabular}[c]{@{}l@{}}proportional to the area of the city\\  relocated.\end{tabular}                                                    \\ \hline
\multicolumn{1}{|l|}{surges}                & \begin{tabular}[c]{@{}l@{}}probability of\\ exceedance\end{tabular}   & annual max surge based on GEV                                                                                                                 \\ \hline
\end{tabular}
\end{table}

\section{Methods}\label{methods}

\subsection{iCOW framework computational description}\label{iCOW framework}

The iCOW framework consists of two modules, the iCOW module and a set of multi-objective evolutionary algorithms (MOEA). Together, these work in tandem to evaluate and optimize multiple risk mitigation strategies.

\subsubsection{iCOW module}\label{iCOW module description}
We developed the iCOW module in C for computational efficiency. It consists of one program with two major components executed in sequence. The first component takes the exogenous factors associated with the initial baseline city and characterizes the city (in terms of the value distribution, zones, and damage functions, described in section \ref{relationships}) in response to the chosen strategy lever settings described in \ref{strategiesAndLevers}. The costs to implement the strategy lever inputs and city value changes are iCOW module output metrics. The second component takes the city's value distribution, zones, and damage functions as inputs, and evaluates the city's response to the exogenous set of storm surge sequences, described in section \ref{surge generation module}, to generate the remaining module output metrics.

For this study, the defensive strategies are established at time zero, and hence, the first component to characterize the city is evaluated only once, and the characteristics for the city are set for all storm surges evaluated. The impacts of storm surge output metrics can include average cost in dollars over a time span (as discussed above), flood frequency, dike breach frequency, or frequency of events over an unacceptable damage threshold. For this study, we use both the cost of implementing strategies and the total cost of damages over 50 years as the input metrics for the MOEA discussed below in sections \ref{MOEA module} and \ref{results}.

\subsubsection{MOEA module}\label{MOEA module}

As more strategies are considered to provide varying degrees of protection, and as the number of objectives increases, selecting optimal solutions becomes much more challenging \citep{hadka_borg:_2013}. Even with a greatly simplified model (compared to CLARA), selecting the `best' mix of surge risk mitigation strategies for a given set of objectives is nontrivial. We use the Borg MOEA to solve the optimization problem because of its high computational efficiency and easy scalability to parallel computing environments \citep{hadka_borg:_2013,hadka_large-scale_2015}. The Borg MOEA is not a single MOEA algorithm, rather, it is an auto-adaptive class of high performance MOEA algorithms based on the evolutionary progress of a population of candidate solutions \citep{hadka_borg:_2013}. We use initial Borg population sizes of 200 members. This initial population evolves over one million functional evaluations to a final larger populations of more than 1,000 members for all figures in this article. We compared results generated using the Borg MOEA with solutions generated by the NSGAII algorithm \citep{goos_fast_2000} and find consistent solutions.

\section{Example Applications}

We explore Pareto optimal strategies for protecting cities against future storm surges. We identify multiple cases with increasing complexity for optimization. We start with the relatively simple `van Dantzig style' single objective, single lever solution (section \ref{single objective}). To demonstrate the capabilities of the iCOW module, we then increase the number of objectives to two and increase complexity by examining the one, two, four, and five lever cases.
 
\subsection{Single objective optimization}\label{single objective}
Optimizing `van Dantzig style' models with one lever (dike height) and one objective (to minimize net present cost) is simple, and in many cases, can be solved analytically \citep[e.g.][]{van_dantzig_economic_1956}. The iCOW model can be configured to emulate `van Dantzig style' behavior by varying implementation of one strategy with the other strategies held constant (i.e. vary dike height, but fix withdrawal height, dike setback, and resistance height to zero), and measuring the net cost that results from every dike height. The optimal solution that emerges is a single level defined by an optimal dike height and a resultant net cost (see supplemental section \ref{van Dantzig} for additional discussion and examples).

Note that a point solution based on lowest net cost may not be realistic. Decision-makers' funds available for storm surge risk mitigation investments might, for instance, compete with funding for mitigation of other risks, or funding for other economically valuable investments (e.g., education or sports facilities.) Thus, fully funding a 'van Dantzig style' economically optimal level may not be possible. Alternatively, other stakeholders may prefer the lower variability and uncertainty in future risk over the more tangible current investment costs. In these cases, being able to examine higher levels of storm surge risk mitigation investment (in terms of reducing the variability of future risk) is appropriate.

\subsection{Multiple objective optimization with multiple levers}\label{multi objective}

To provide decisionmakers with more options, we increase the complexity of the problem by incrementally increasing the number of levers to provide many combinations of strategies and determine Pareto optimal two objectives solutions. We identify three cases to help illustrate the additional complexities when optimizing across multiple strategies simultaneously.  For the first case, we compare the cost and risk reduction associated with a dike only single lever (\textbf{D}) strategy, with a dike height (\textbf{D}) and resistance height (\textbf{R}) two lever strategy. In the single lever example, a dike can be constructed at the seawall to any height. The resultant city consists of zones 3 and 4. In the two lever example, policy makers can either construct a dike or incorporate resistance (with a fixed percentage, P). Given the structure of the iCOW model, the two levers are mutually exclusive. The resultant city then consists of either zones 3 and 4, or zones 2 and 4. 

For the second case, we add two additional levers, dike base location (\textbf{B}) and resistance percentage (\textbf{P}) and compare results to the first case. In this four level example, dike height and resistance height are no longer mutually exclusive. Dikes can be constructed to any height at any location, and resistance can be incorporated in front of the dike to any height at any percentage. 

For the third illustrative case, we add the additional lever of complete withdrawal from the lowest city levels, (a five lever example) and compare results with the other cases. 

\section{Results}\label{results}

\begin{figure}[h]
\begin{flushleft}
\includegraphics[width=5in]{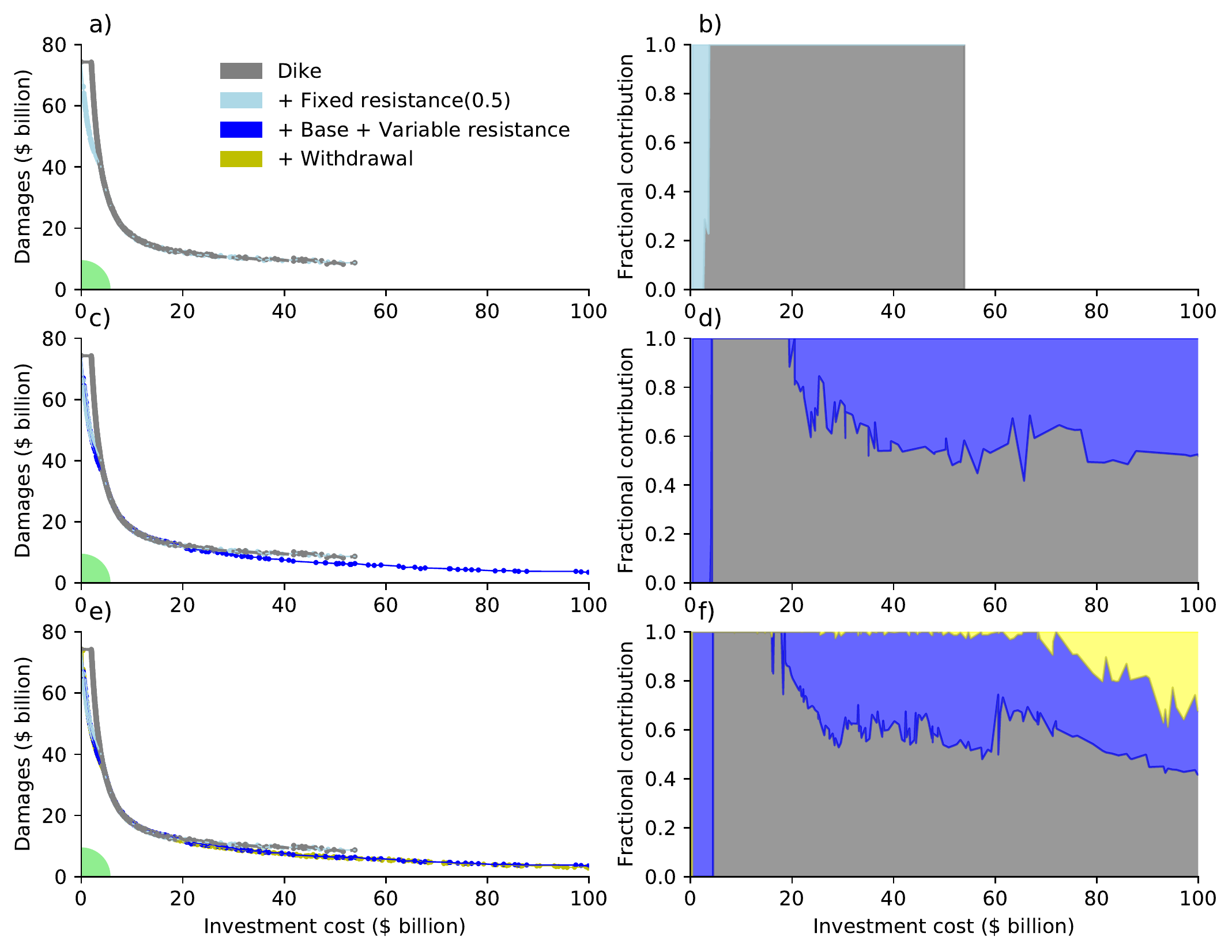}
\caption[iCOW evolution of the Pareto front]{Optimal two objective (damage and investment costs) risk mitigation strategies identified by the iCOW framework. Panel a and b shows the one and two lever cases,  panel c and d show the four lever case, and panel e and f show the five lever case. The resultant Pareto fronts are shown in the left panels (a, c, and e). The right panels (b, d, and f) are normalized stacked area charts that show the fractional contributions of individual strategies (indicated by color on the vertical axis) for the corresponding total investment cost (horizontal axis). The green quarter circles in panels a, c, and e indicate the location of desired solutions that have both low investment cost and low damages. }
\label{fig3 Pareto Fronts}
\end{flushleft}
\end{figure}

For the first case, we consider when the only solution available is construction of a dike. In examining this case with one and two levers (see Fig.~\ref{fig3 Pareto Fronts}a),  the iCOW framework identifies an economic barrier to entry situation, when no optimal solutions emerge as investment level increases until the investment cost exceeds the dike fixed start-up cost. In this situtation, when the dike is so low, the increased damages associated with a failed or overtopped dike exceeds the damages avoided when the dike is not present. Once this dike investment threshold is reached the framework continues to identify solutions with taller, more expensive dikes until additional dike height no longer reduces damages. Incorporating a resistance option allows policy makers to identify sensible low cost solutions that substantially decrease storm surge damage. However, once the dike investment threshold is reached, the optimal strategy shifts abruptly to increasing dike height. This can be seen in Fig.~\ref{fig3 Pareto Fronts} panel b. Before the transition point, 100\% of investment cost is applied to the resistance strategy, while after the transition, 100\% of investment cost is applied to a dike strategy. 

%%%
%%%

\clearpage

For the second case (with four levers), the optimal strategy combinations that emerge are superior to the one and two lever examples in that equal or lower damages can be obtained for any given investment level (see Fig.~\ref{fig3 Pareto Fronts}c). Fig.~\ref{fig3 Pareto Fronts}d shows the shift in the percentage of optimal investment for any given level of investment allocated to each strategy that occurs as the level of investment increases. At low levels of investment, the optimal strategy is based solely on implementation of resistance. At higher investment levels, better performance is achieved using a dike only solution. At the highest considered levels of investment, the optimal solution consists of combining strategies by setting the dike back from the seawall and implementing resistance in the area between the waterfront and dike base.

In the third case (with five levers), the additional withdrawal lever allows for the identification of further Pareto improvements at the highest investment levels, as shown in Fig.~\ref{fig3 Pareto Fronts} e and f).

In both the four and five lever cases, jagged fractional contribution are evident between \$30 - \$60 billion in Fig.~\ref{fig3 Pareto Fronts} panels d and f. These variations occur when when the BORG MOEA algorithm identifies diverse solutions with approximately equivalent trade offs between the two objectives (damage and investment cost) resulting over multiple combinations of lever settings.

\begin{figure}[h]
    \begin{flushleft}
    \includegraphics[width=5.4in]{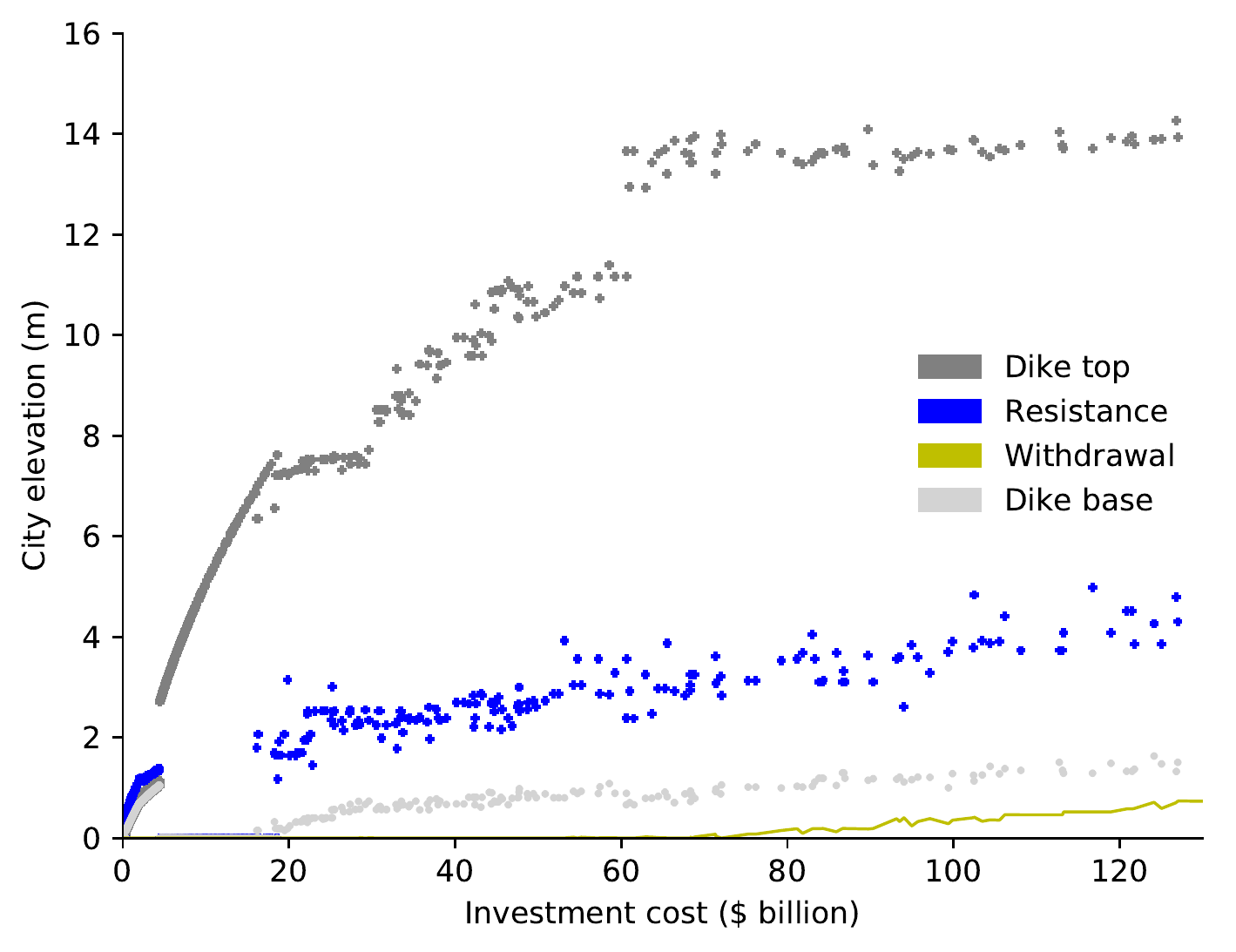}
    \caption[Strategy heights profile]{Dependence of city elevation heights on investment costs derived from the set of Pareto optimal risk mitigation strategies using five levers and two objectives.}
    \label{Strategy City Elevation Profile}
    \end{flushleft}
\end{figure}

To help visualize the solutions for a given level of investment, we can also identify the city elevations for the Pareto optimal strategies (Fig.~\ref{Strategy City Elevation Profile} for the five lever example). At the lowest levels of investment, the lowest cost damage reduction strategy is to implement resistance into an increasing area of the city. Once a sharp cost threshold is reached, the Pareto optimal strategy shifts abruptly to implementation of an increasingly tall dike at the city seawall. As the amount of investment continues to increase, corresponding to the objective to further decrease damage, the optimal solution sets the dike back from the seawall at gradually increasing elevations and with slightly increasing dike heights. The area between the seawall and the dike is fortified through implementation of resistance first and then as investment continues to increase, through a combination of withdrawal and resistance. The jagged fractional contributions visible in Fig.~\ref{fig3 Pareto Fronts} panel f manifests itself in Fig.~\ref{Strategy City Elevation Profile} as variations in the city elevations for adjacent (in terms of investment cost) strategies.

The character of these optimal solutions are highly dependent and sensitive to iCOW model parameters. For instance, very small parameter changes to the cost of resistance relative to withdrawal and dike costs can result in surprising and nonlinear changes to the character of the resulting solution sets. 

%%%
%%%
\clearpage

\section{Discussion and limitations}\label{discussion}
Simple conceptual storm surge damage models may fail to provide sufficient fidelity in simulating storm surge risk mitigation. More complicated regional storm surge damage models improve upon these simple models in terms of realism, fidelity, and the area covered, but they can be difficult to design and implement. Additionally, they are computationally expensive to evaluate multiple combinations of risk mitigation strategies over a wide range of future storm surge risks, while considering the differing objectives of various stakeholder communities. The combination of these drawbacks may make this approach impracticable for some communities.

We use the iCOW framework to analyze an idealized city designed to resemble major cities such as New York. Every coastal community, however, is unique, thus many iCOW assumptions (such as the assumptions regarding the regular wedge shape, the uniform density, or the high value density of the island city) and parameters (such as the costs to implement resistance or the damage functions associated with different zones) will not always be appropriate. Moreover, local policymakers are likely to have better local information available as to the proper values for many iCOW parameters. For instance, local experience in improving buildings to make them more damage resistant will provide cost information that better reflects local circumstances.  To make iCOW useful to city planners at a specific site, we must adjust iCOW to reflect city specific characteristics. At a minimum, such site specific adjustments need to account for the following factors: site specific projections of storm surges, local economic demographics and their relationship to the site topography, typical building heights, and adjustments to damage functions that account for site specific building usage and construction type. Additionally, some simplifying assumptions described in this paper would probably need to be refined. Examples of these could include the assumption of uniform city density and value with respect to elevation from the waterfront, the cost basis for constructing dikes or implementing resistance, or the simplifying assumption that a single surge level is the best factor relating projected storm characteristics to damage.

Two objective optimal solutions are sensitive to iCOW parameters as discussed above, but fully evaluating this sensitivity has not yet been accomplished and may be computationally infeasible. Conducting sensitivity analysis with more than two objectives will likely be even more challenging. In the course of adapting iCOW to a particular community, a sensitivity analysis would be centered on the particular combination of assumptions and parameters used in simulating the community under study to provide insight for decisionmakers into critical uncertainties and to focus limited resources towards understanding and mitigating those uncertainties.

\section{Conclusions}\label{conclusions}

We develop the iCOW as a storm surge risk modeling framework of intermediate complexity for decisionmakers and stakeholders. The framework is intended to fill the gap between simple conceptual storm surge damage models and realistic, but complex and expensive, state-of-the-art regional storm surge risk modeling frameworks. We demonstrate the capabilities of the iCOW framework to evaluate and optimize surge risk mitigation strategies ranging from a simple single-objective, single-lever problem to a more complicated and computationally challenging two-objective five-lever problem. Decisionmakers can use iCOW to explore a more comprehensive set of strategies over a wider range of future risk scenarios than previous done with more complicated regional state-of-the-art models. Insights gained by this approach, however, will be more limited in geospatial extent (to a single area of fairly uniform characteristics) and in terms of less realistic representation of specific future outcomes. Stakeholders can use iCOW to supplement insights gained from those higher complexity models and illustrate trade offs between conflicting objectives. iCOW provides one degree of spatial resolution in terms of distance from the waterfront, and thus represents a compromise between the point (single polder) solutions provided by `van Dantzig style' approaches, and regional gridded results produced by state-of-the-art frameworks. iCOW is relatively simple to modify and implement, and so it may be useful to decisionmakers with limited resources, but who nevertheless need methods to identify, evaluate, and illustrate the multiple trade offs implicit in any storm surge risk mitigation strategy. The iCOW framework can be easily modified or extended and can be integrated with other modeling systems such as the Building blocks for Relevant Ice and
Climate Knowledge (BRICK) modeling framework \citep{wong_brick_2017} to further explore storm surge risk in a regional or global context.

\section{Acknowledgements}\label{ack}

We thank D. Hadka for outstanding technical support with the Rhodium multi objective tool kit and the Borg MOEA. We thank B. Lee, M. Haran, D. Titley, R. Lempert, and J. Lawrence for helpful discussions. This research was partially supported by the National Science Foundation (NSF) through the Network for Sustainable Climate Risk Management (SCRiM) under NSF cooperative agreement GEO-1240507 and the Penn State Center for Climate Risk Management. Any opinions, findings, and conclusions or recommendations expressed in this material are those of the authors and do not necessarily reflect the views of the NSF. Errors and opinions are, of course, those of the authors.

All authors contributed to the iCOW conceptual design. RC developed detailed model designs, wrote all software code and integrated iCOW software with the BORG MOEA algorithm. All authors developed the experimental plan. RC designed and executed experiments and conducted analysis of results. CF and KK provided guidance for the project. RC wrote the article and all authors contributed to editing.

\section{Code availability}
All iCOW software code used to create the figures in this article are currently available from the corresponding author, they will be open source under a GNU non commercial license and distributed by a repository upon publication as a peer reviewed study. 

\section{Supplemental material}
Supplemental material in the appendices provides additional discussion on emulation of van Dantzig style model emulation, explanations and implications of using GEV based storm surges, and information on convergence and computational expense associated with changing the number of futures evaluated.

\section{Conflict of interest}
The authors are not aware of financial or personal relationships that would pose a conflict of interest.
%%\end{linenumbers}
\clearpage

\appendix

\section{`van Dantzig style' model emulation}\label{van Dantzig}

\begin{figure}[h]
    \begin{flushleft}
    \includegraphics[width=5.2in]{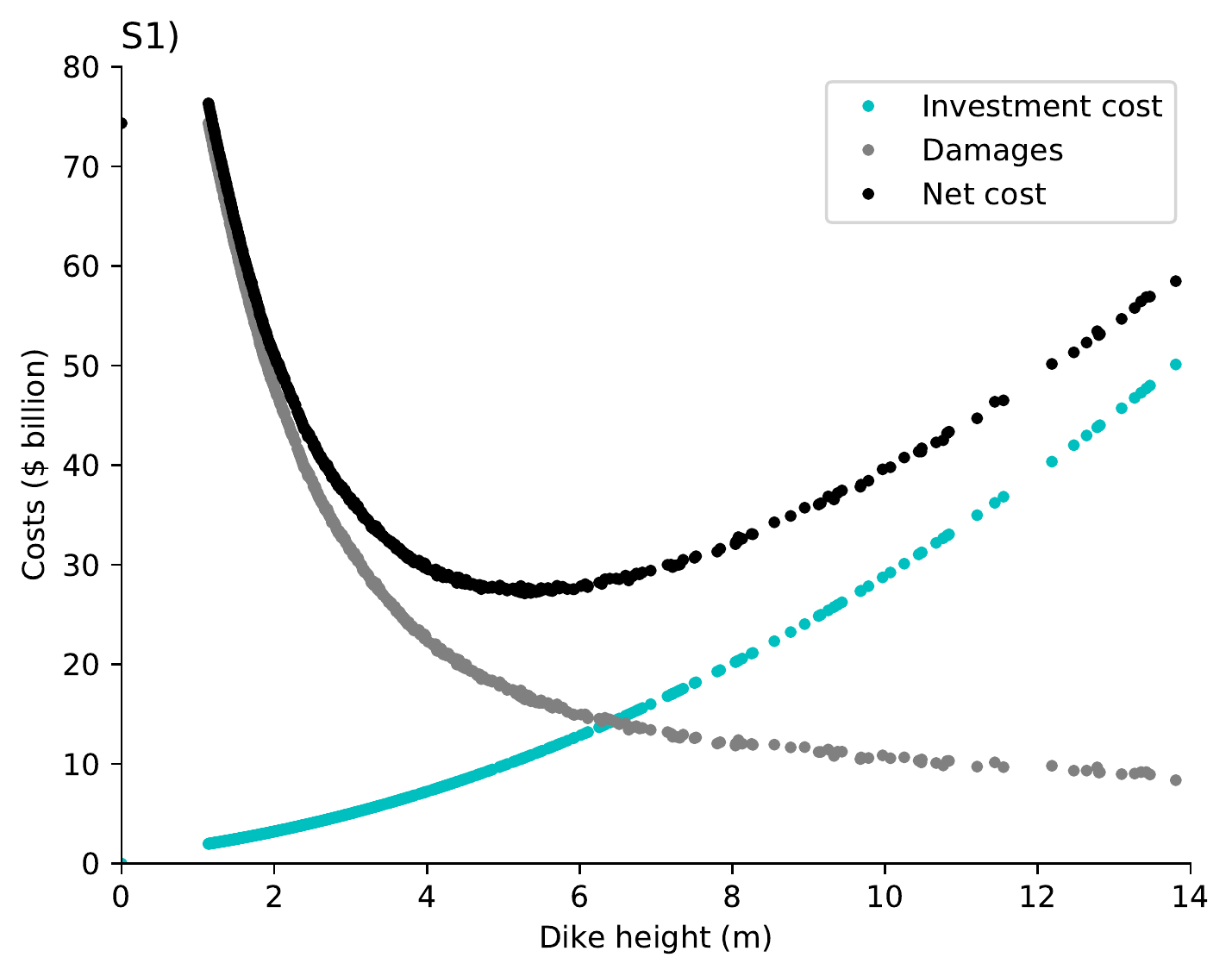}
    \caption[van Dantzig iCOW emulation]{iCOW emulation of `van Dantzig style' optimal dike height. The optimal dike height occurs when the net cost is minimized, equating to a dike height of approximately 5.5m.}
    \label{fig_van_Dantzig}
    \end{flushleft}
\end{figure}

The iCOW framework is capable of emulating `van Dantzig style' single objective optimization. For example, the `van Dantzig style' optimal dike height for the illustrative city used in this journal is shown in Fig.~\ref{fig_van_Dantzig}. While the emulation produces a single `optimal' dike height and a visually similar result, there are some differences and the resultant optimal dike heights will not be equal to those obtained directly from \cite{van_dantzig_economic_1956}.  The iCOW `van Dantzig' emulation shown in Fig.~\ref{fig_van_Dantzig} shows optimization of dike height with respect to the net cost over a fixed 50 year timespan. The optimization performed by van Dantzig did not consider a fixed time interval \citep{van_dantzig_economic_1956}. His solution comprised an optimal height based on then current analysis of tidal records and considering the net present value of both current investments and damages incurred for the entire span of the future. \citep{van_dantzig_economic_1956}.  The dike cost model used in \cite{van_dantzig_economic_1956} is a simple linear relationship to dike height that incorporates no start up costs. iCOW dike cost is based on the wedge-shaped geography of the City Model, the resultant dike volume, and a startup cost as described in section \ref{DikeRelationships} and equations \eqref{dike volume} and \eqref{dike cost}. The resultant relationship between dike height and cost is nonlinear.

%%%
%%%
Fig.~\ref{fig_van_Dantzig} shows a gap between a zero dike height and approximately 0.7 m where no solutions exist. Moreover, Fig.~\ref{fig_van_Dantzig} shows an increase in net cost between the no investment maximum damage point and next optimal point above this gap. This gap and subsequent increase in net cost is due to the fixed start up cost of dikes assumed in the iCOW module that does not occur in van Dantzig's model \citep{van_dantzig_economic_1956}, for instance, as implemented by \cite{oddo_deep_2017}. Additionally, because the iCOW framework increases damage behind a dike if the dike fails or is breached, at very low dike heights, total damage increases with increased dike height. The BORG MOEA algorithm therefore does not identify any optimal solutions in these dike height ranges. See discussion in \ref{DikeRelationships}.

\clearpage

\section{Relationship between states of the world evaluated and wall time}\label{futures}
\begin{figure}[h]
    \begin{flushleft}
    \includegraphics[width=5.4in]{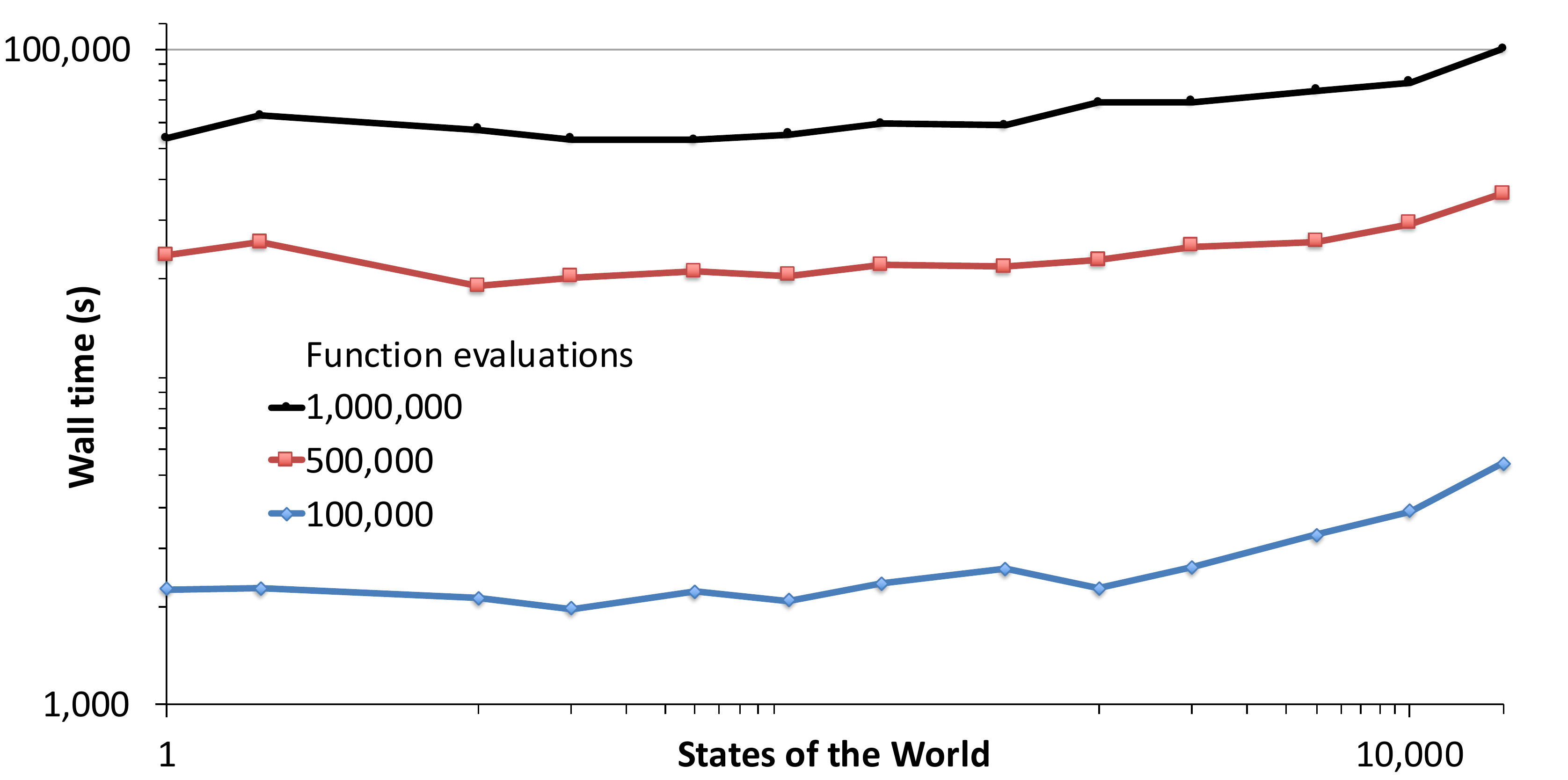}
    \caption[iCOW computational cost]{Relationship between computational cost and states of the world evaluated. The wall time in seconds (y-axis) required to evaluate states of the world (where one state of the world is a 50 year sequence of annual highest storm surges) (x-axis)}
    \label{Computational cost futures}
    \end{flushleft}
\end{figure}

Each iCOW framework state of the world consists of an exogenous 50-year sequence of storm surges. The representation of risk associated with the most extreme storm surges improves with more states of the world evaluated. Evaluating more states of the world, however, adds computational cost. To understand this relationship, we measure the computational wall time required to conduct five lever iCOW optimizations using 1 to 20,000 states of the world for 100,000, 500,000, and 1 million functional evaluations (Fig.~\ref{Computational cost futures}). We select 5,000 states of the world as a compromise between computational efficiency and stable results for use in all experiments described in this article.

%%%
%%%
\clearpage

\section{iCOW model parameters}\label{parameter tables}

\begin{table}\label{tabInitialCityParameters}
\caption{iCOW parameters}
\begin{tabular}{|l|r|l|l|}
\hline
Name  & Value & Unit & Description    \\ 
\hline
$v_i$   & 1.5   & \$T   & initial city value \\ \hline
$B$   & 30    & m   & building height\\
\hline
$H_{city}$       & 17    & m  & height of the city \\
\hline
$Depth_{city}$       & 2     & km  & distance from seawall to highest point\\
 \hline
$Length_{city}$       & 43    & km  & length of the lowest (seawall) coast \\
\hline
$H_{seawall}$    & 1.75  & m & seawall height   \\ \hline
%%\end{tabular}
%%\label{tabInitialCityParameters}
%%\end{table}

%%\begin{table}
%%\caption{iCOW cost algorithm parameters}
%%\begin{tabular}{|l|r|l|l|}
%%\hline
%%Parameter   & Value & Units  & Description                                  \\ \hline
Dike        & 3   & m   & equivalent height added to actual height     \\ 
start       &     &     & so larger dikes have larger startup costs.   \\ \hline
Dike top    & 4   & m   & controls the width of the dike top to   \\ 
width       &     &     & adjust dike volume.     \\ \hline
$s$        & 0.5 & m   & slope of the dike sides. \\ 
\hline
Dike cost   & 10 & \$/m$^3$   & dike cost per volume  \\ \hline
Dike value  & 1.1 & n/a    & controls increase in value for areas\\ 
ratio       &     &     & protected by a dik                                        \\ \hline
unprotected & 1.0 & n/a    & controls increase in value for areas\\
ratio       &     &     &  protected                                              \\ \hline
$f_w$ & 1.0 & none & withdrawal factor adjusting cost to relocate \\ \hline
Withdrawal  & 0.01 & n/a & fraction of population and infrastructure\\
fraction    &     &     & that will leave if withdrawn                       \\ \hline
$f_{lin}$       & 0.35    & n/a    & Linear factor relating percent resistant to \\
 & & & resistance cost. \\ \hline
$f_{exp}$        & 0.9 & n/a    & Factor for exponential cost increase when\\
 & & & percent resistant > $t_{exp}$. \\
\hline
$t_{exp}$        & 0.6 & n/a & percent resistant threshold above which\\
 & & & resistance cost increases exponentially. \\
\hline
 
Basement   & 3.0   & m  & representative building height of damage\\ 
depth      &       &    & occurring when surge reaches building.\\ 
\hline
Futures   & 5000   & n/a & number of 50 year storm surge sequences\\ 
evaluated      &       &    & used for each BORG functional evaluation.\\ 
\hline
Borg   & 5000   & \$K  & BORG epsilon for 1 and 2 levers \\ 
epsilon      & 500      &  \$K  & BORG epsilon for 4 and 5 levers\\
\hline
Pop_{initial}& 200 & n/a & initial Borg number of iCOW cities.\\
\hline
nfe & 100 & K & number of Borg functional evaluations.\\
\hline
\end{tabular}
\label{costParemeters}
\end{table}

\begin{table}
\caption{iCOW damage algorithm parameters}
\begin{tabular}{|l|r|l|l|}
\hline
Parameter   & Value & Units  & Description                                  \\ \hline
$f_{damage}$     & 0.39  & none & fraction of inundated buildings \\
                 &       &      & damaged. \\  \hline
Protected        & 1.3   &  none  & increased damage that occurs in \\ 
damage factor       &     &     & protected areas if dike fails.   \\ \hline
$t_{df}$    & 0.95   & none   & dike height fraction above which failure   \\
   &     &     & probability increases with surge height.     \\ \hline

Threshold     &1/375 & none   & fraction of city value above which  \\ 
level       &     &    & damage increases more rapidly.          \\ \hline
Wave runup  & 1.1 & none    & surge factor when surge exceeds seawall   \\ 
            &     &         & height.   \\  \hline
\end{tabular}
\label{damageParemeters}
\end{table}



\clearpage

\section*{References}

%%\bibliographystyle{elsarticle-harv.bst} 
%%\bibliography{ICOW.bib}


\end{document}